\documentclass[twocolumn,floatfix,showpacs,preprintnumbers,superscriptaddress,pre]{revtex4-1}
\usepackage{graphicx}
\usepackage{color}
\usepackage{amsmath,amssymb}
\usepackage{revsymb4-1}
\usepackage{bm}

\begin{document}
\title{Momentum and Mass Fluxes in a Gas Confined between Periodically Structured Surfaces at Different Temperatures}

\author{Alexander A. Donkov}
\email[]{aadonkov@csi.tu-darmsdadt.de, aadonkov@uwalumni.com}
\affiliation{Center of Smart Interfaces, Technische Universit\"{a}t Darmstadt, D-64287 Darmstadt, Germany}
\author{Sudarshan Tiwari}
\email[]{sudarshan.tiwari@itwm.fraunhofer.de}
\affiliation{Fachbereich Mathematik,Technische Universit\"at Kaiserslautern, D-67663 Kaiserslautern, Germany}
\author{Tengfei Liang}
\email[]{tfliang@ust.hk}
\affiliation{Department of Mechanical Engineering, Hong Kong University of Science and Technology, Clear Water Bay, Kowloon, Hong Kong}
\author{Steffen Hardt}
\homepage[]{http://www.csi.tu-darmstadt.de/institute/nmf/}
\email[]{hardt@csi.tu-darmstadt.de}
\affiliation{Center of Smart Interfaces, Technische Universit\"{a}t Darmstadt, D-64287 Darmstadt, Germany}
\author{Axel Klar}
\email[]{klar@mathematik.uni-kl.de}
\affiliation{Fachbereich Mathematik,Technische Universit\"at Kaiserslautern, D-67663 Kaiserslautern, Germany}
\author{Wenjing Ye}
\email[]{mewye@ust.hk}
\affiliation{Department of Mechanical Engineering, Hong Kong University of Science and Technology, Clear Water Bay, Kowloon, Hong Kong}
\date{\today}

\begin{abstract}
It is well known that in a gas-filled duct or channel along which a temperature gradient is applied, a thermal creep flow is created. Here we show that a mass and momentum flux can also be induced in a gas confined between two parallel structured surfaces at different temperatures, i.e. \textit{orthogonal} to the temperature gradient. We use both analytical and numerical methods to compute the resulting fluxes. The momentum flux assumes its maximum value in the free-molecular flow regime, the (normalized) mass flux in the transition flow regime. The discovered phenomena could find applications in novel methods for energy-conversion and thermal pumping of gases.
\end{abstract}


\maketitle

\section{Introduction} 

Thermally induced gas flow and Knudsen forces arising from the unbalanced momentum flux in a gas are among the most interesting phenomena in a rarefied gas with a non-uniform temperature field. The rapidly developing micromachining technologies have made it possible to observe and apply these phenomena even at atmospheric conditions. For example, attractive and repulsive forces have been observed on heated microcantilevers in air~\cite{Passian:2003fk,Gotsmann:2005fk}, and a passive micromachined gas pump utilizing thermal creep flow, entitled as the Knudsen pump, has been demonstrated~\cite{McNamara:2005uq}. The promising potential of these applications has called for a deeper understanding and study of transport phenomena in a rarefied gas. Several experimental and numerical studies have been conducted to study the Knudsen force on cantilevers~\cite{Passian:2002kx,Passian:2003qf,Lereu:2004ys,Zhu:2010zr}, radiometric forces~\cite{Selden:2009ly}, and Knudsen pumps~\cite{Muntz:2002ve}.

Most Knudsen pumps proposed and developed up to now utilize thermal creep flow. The modeling of such flows dates back to 1879 when Maxwell~\cite{Maxwell:1879fk} derived a mathematical expression for the flow along a surface induced by a temperature gradient. Since then, the corresponding thermal creep flow through ducts or channels has been demonstrated in many experiments. The flow velocity is proportional to the temperature gradient, potentially requiring very large temperature spans when the channel length is no longer small, which has limited the application perspectives of thermal creep flow. Some attempts have been made to reduce the required temperature span by considering a gas between parallel surfaces with a periodic, step-like topography~\cite{Sone:1996fk}. Besides that, virtually no studies of transport phenomena in gases confined between two periodically structured surfaces seem to be available.

\begin{figure}[b!]
\begin{center}
\includegraphics[width=3in]{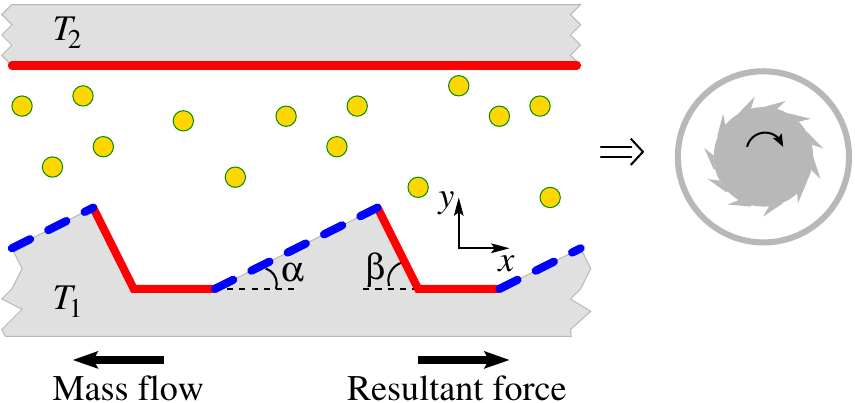}
\caption{\label{fig:device} (Color online) Schematic of the two-dimensional channel geometry considered made of two surfaces with temperatures $T_1$ and $T_2$. Different styles of the wall segments indicate different boundary conditions. On the right side, a corresponding geometry incorporating a wheel with suitable surface structures is sketched to suggest possible applications. }
\end{center}
\end{figure}

It is well known that in a gas confined between two parallel isothermal walls of different temperatures no net mass and momentum fluxes parallel to the walls occur. However, as we will show in this article, with periodically structured surfaces it is possible to induce such fluxes. We consider a monoatomic gas governed by the Boltzmann equation between two parallel surfaces equipped with periodic structures. The term "structure" has a twofold meaning. On the one hand, the surfaces can have a topography, on the other hand, there can be spatial modulations of the boundary condition expressing how a molecule gets reflected when impinging on a wall. We expect that especially when the Knudsen number $\text{Kn}$, being the ratio of the mean-free path of the molecules and the distance between the surfaces, is of the order of one or larger, the surface topography and the spatial modulation of the boundary condition should have a significant influence on the phase space distribution of the molecules inside the channel and the resulting transport phenomena. We are mainly interested in the total momentum and mass fluxes inside the gas that are created if the two surfaces are at different temperatures. The purpose of this work is to identify surface structures that are as simple as possible and nevertheless give rise to non-vanishing fluxes. The type of prototype geometry considered is depicted in Fig.~\ref{fig:device} together with the predicted directions of mass and momentum fluxes. The quantities of interest are the total momentum and mass fluxes in $x$-direction. It is clear that a non-vanishing flux along the $x$-direction will only result if the structure breaks the reflection symmetry with respect to $x$. As will become evident in the following, it suffices to structure one of the two surfaces. The top wall will therefore be considered flat and uniform.

\section{Momentum flux} 

\begin{figure}[t!]
\begin{center}
\includegraphics[width=3in]{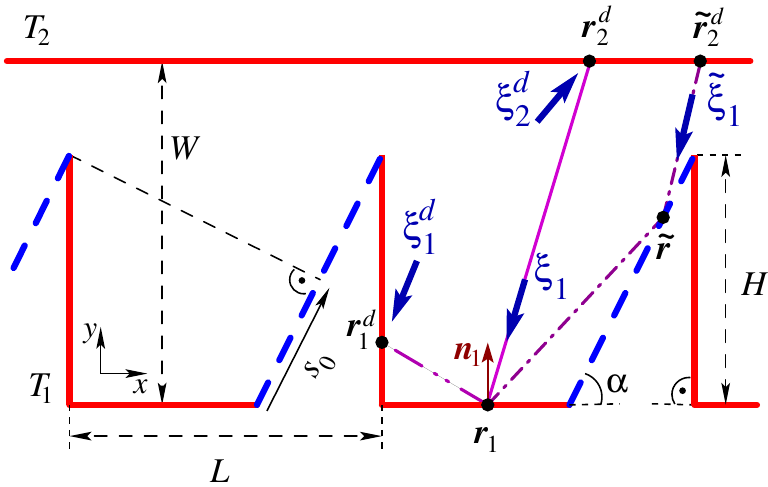}
\caption{\label{fig:manyfinsgeometry-notations} (Color online) The model geometry. Shown are the notations and an example of a tracing starting at $\bm{r}_1.$}
\end{center}
\end{figure}

Specifically, we are interested in the total momentum in $x$-direction per unit time the gas transfers between the top and the bottom surface. To further simplify matters, let us limit ourselves to the surface structures shown in Fig.~\ref
{fig:manyfinsgeometry-notations}. Besides choosing $ \beta = \pi/2$, the distribution of wall boundary conditions has been fixed. The red (solid) sections denote a diffuse reflection boundary condition, the blue (dashed) ones a specular reflection boundary condition, whose mathematical form will be detailed below. Furthermore, for the time being we make the assumption that the mean-free path of the molecules is much larger than $W$, i.e. we assume gas dynamics in the free-molecular flow regime. We always suppose that the thermal conductivity of the wall material is so large that temperature variations along the surfaces can be neglected. 

Assuming no external force fields acting on the molecules, in the limit $\text{Kn} \to \infty$ the solution of the Boltzmann equation can be reduced to tracing rays representing the paths of the molecules~\cite{Sone:2007,Cercignani:2000,Kogan:1969}. 
We give a sketch of the derivation based on the approach presented in ref.~\cite{Kogan:1969}.

We denote by $\bm{r}_1$ a point on the lower surface and by $f_i(\bm{r}_1, \bm{\xi}_1)$ the distribution function of the incoming molecules incident with velocity $\bm{\xi}_1$ on $\bm{r}_1.$ For those we have  $\left(\bm{\xi}_1 \cdot \bm{n}_1 \right)<0,$ where $\bm{n}_1$ is inward normal at $\bm{r}_1$ (in this and the following 
the notation for the unit normals is analogous to the boundary points). The force exerted on the bottom surface by the incoming molecules is then:
\begin{equation}
\bm{F}_i = -\int  \int_{\bm{\xi}_1 \cdot \bm{n}_1<0}m \bm{\xi}_1 (\bm{\xi}_1 \cdot \bm{n}_1)  f_i(\bm{r}_1(s_1),\bm{\xi}_1) \textrm{d}^2\bm{\xi}_1ds_1,
\label{eq:eqForce_in}
\end{equation}
where the integration is over an arc length parameter $s_1$ through which the points at the bottom wall are defined via $\bm{r}_1(s_1)$.
Similarly, the recoil from the outgoing molecules at $\bm{r}_1$ gives:
\begin{equation}
\bm{F}_r =- \int  \int_{\bm{\xi}_1 \cdot \bm{n}_1>0} m \bm{\xi}_1 (\bm{\xi}_1 \cdot \bm{n}_1) f_r(\bm{r}_1(s_1), \bm{\xi}_1) d^2\bm{\xi}_1 ds_1,
\label{eq:eqForce_out}
\end{equation}
where $f_r(\bm{r}_1,\bm{\xi}_1)$ is the phase space distribution of the outgoing molecules. 
The net force on the lower surface is then given by $ \bm{F}=\bm{F}_i+ \bm{F}_r.$

In the framework of the collisionless Boltzmann equation $f_i(\bm{r}_1, \bm{\xi}_1)$ is equal to $f_r(\bm{r}, \bm{\xi}_1),$ where $\bm{r}$ is a point at the intersection of a straight line oriented along $-\bm{\xi}_1$ with the boundary.  The point $\bm{r}$ could be either on a diffusive or on a specular boundary, a situation depicted in Fig.~\ref{fig:manyfinsgeometry-notations} with the possibilities for $\bm{r}$ exemplified by $\bm{r}^d_1,$ $\bm{r}^d_2$ for the diffusive and by $\bm{\tilde{r}}$ for the specular wall segments, respectively.

The boundary condition on the wall segments with diffuse reflection of molecules is expressed as
\begin{multline}
f_r(\bm{r}^d, \bm{\xi}_1) = \nu(\bm{r}^d) \mathsf{F}^{2 D}(\bm{r}^d,\xi_1),\\
 \mathsf{F}^{2 D}(\bm{r}^d,\xi_1) =\frac{2}{\sqrt{\pi}}\left( \frac{m}{2 T_{\bm{r}^d}} \right)^{3/2} \exp\left(-\frac{m \bm{\xi}_1^2}{2 T_{\bm{r}^d}} \right),
\label{eq:f_diffusive}
\end{multline}
where $m$ is the molecular mass, $T_{\bm{r}^d}$ the temperature of the wall at point $\bm{r}^d$ (in energy units), and the number of collisions from the incoming molecules at $\bm{r}^d$ per unit length and unit time (the particle flux) $\nu(\bm{r}^d)$ has been factored out. 
This is a Maxwell boundary condition~\cite{Maxwell:1879fk} with an accommodation coefficient of one. 
The particle flux is defined in the usual way as:
$\nu(\bm{r}^d) = -\int_{\bm{\xi}^d \cdot \bm{n}^d < 0} \left(\bm{\xi}^d \cdot \bm{n}^d \right) f_i(\bm{r}^d,\bm{\xi}^d) \textrm{d}^2\bm{\xi}^d.$ 

From particle number conservation we have $\nu(\bm{r}^d)= \int_{\bm{\xi}_1 \cdot \bm{n}^d > 0} \left(\bm{\xi}_1 \cdot \bm{n}^d \right) f_r(\bm{r}^d,\bm{\xi}_1) \textrm{d}^2\bm{\xi}_1,$ 
from which the choice of the pre-exponential factor in Eq.~\eqref{eq:f_diffusive} is fixed.

For the specular boundary segments the corresponding boundary condition is $f_r(\bm{\tilde{r}}, \bm{\xi}_1) = f_i(\bm{\tilde{r}}, \bm{\tilde{\xi}}_1),$
where the expression on the right-hand side is obtained by further tracing in the direction of $-\bm{\tilde{\xi}}_1.$
The velocity $\bm{\tilde{\xi}}_1$ is related to $\bm{\xi}_1$ by reflection from the surface with a normal vector $\bm{\tilde{n}}$, $\bm{\tilde{\xi}}_1=\bm{\xi}_1-2(\bm{\tilde{n}}\cdot \bm{\xi}_1)\bm{\tilde{n}}.$ The specular boundary condition is again Maxwell's boundary condition, but for a wall with an accommodation coefficient of zero. Tracing particle paths backwards from specular wall segments always brings us to points on a diffuse boundary, in our example in Fig.~\ref{fig:manyfinsgeometry-notations} the point $\bm{\tilde{r}}^d_2$ at the upper wall, where the distribution function of reflected molecules is given by Eq.~\eqref{eq:f_diffusive}. By similar tracing procedures we find the relevant relations for the distribution function in the equation for the recoil force Eq.~\eqref{eq:eqForce_out}. 

As a result, finding the phase space distribution of the molecules at points along the boundary of the domain essentially reduces to determining $\nu(\bm{r})$. 
For illustration, a point $\bm{r}_1$ at the bottom wall receives particles traced to either a diffusive or to a specular wall segment. 
The particle flux's definition then gives an integral equation
\begin{multline}
\nu(\bm{r}_1) =- \int\limits_{\Omega^d(\bm{r}_1)} (\bm{\xi}_1 \cdot \bm{n}_1) \nu(\bm{r}^d) \mathsf{F}^{2 D}\bigl(\bm{r}^d, \bm{\xi}_1\bigr) \textrm{d}^2\bm{\xi}_1\\
- \int\limits_{\tilde{\Omega}(\bm{r}_1)} (\bm{\xi}_1 \cdot \bm{n}_1) \nu(\bm{\tilde{r}}^d) \mathsf{F}^{2 D}\bigl(\bm{\tilde{r}}^d, \bm{\tilde{\xi}}_1\bigr) \textrm{d}^2\bm{\xi}_1,
\end{multline}
where  $\Omega^d(\bm{r}_1)$ denotes the subset of incoming velocities $\bm{\xi}_1$ for which the tracing routine relates the point  $\bm{r}_1$
to a point on a diffusive segment $\bm{r}^d= \bm{r}^d\bigl(\bm{r}_1,\bm{\xi}_1\bigr),$ and similarly,  
$\tilde{\Omega}(\bm{r}_1)$ denotes the subset of incoming velocities for which the tracing routine relates the $\bm{r}_1$ to a point on the specular segment $\bm{\tilde{r}},$ and eventually to a diffusive boundary point $\bm{\tilde{r}}^d= \bm{\tilde{r}}^d\bigl(\bm{r}_1,\bm{\xi}_1\bigr).$ By direct substitution we find that this integral equation has a constant solution, i.e. $\nu(\bm{r})=\nu$, independent of position.

To proceed, let us denote by $s_0$ the projection of the tip of a fin onto the incline of its neighboring fin, cf.~Fig.~\ref{fig:manyfinsgeometry-notations}. For close-by fins or for sufficiently large values of $\alpha,$ $s_0$ is positive, otherwise negative. 
This parameter allows to conveniently distinguish geometries such that a tracing procedure starting at $\bm{r}_1$ and hitting the incline can only end at points on the top wall ($s_0<0$), and geometries for which the tracing after hitting the incline can end either on the top wall or on the vertical side wall of the lower surface ($s_0>0$).
Evaluating the integrals in Eqs.~\eqref{eq:eqForce_in} and~\eqref{eq:eqForce_out} yields  for one unit cell of length $L:$
\begin{equation}
F_x =
\left\{ 
\begin{split}
&A \sin(\alpha)\Bigl(\frac{\pi}{2} - \alpha\Bigr) L \cos(\alpha), \quad \quad \quad \quad (s_0>0)\\
&A\sin(\alpha) \left[\Bigl(\frac{\pi}{2} - \alpha \Bigr) L \cos(\alpha)-\right. \\
&\left.\quad \quad \quad \quad - s_0 \arctan \left(\frac{s_0}{L \sin (\alpha)}\right)\right] , (s_0<0)
\end{split}
\right.
\label{eq:fx}
\end{equation}
where $A=\sqrt{\frac{2 m}{\pi}} \nu \bigl(\sqrt{T_2}-\sqrt{T_1}\bigr),$
and
\begin{equation}
F_y = - \sqrt{\frac{2 m}{\pi}} \nu  \bigl( \sqrt{T_1} + \sqrt{T_2} \bigr) \frac{\pi L}{2}- \frac{F_x}{\tan(\alpha)}.
\label{eq:fy^top}
\end{equation}

In the limit $L \rightarrow \infty$ we have $F_x \propto \Bigl[\frac{\pi}{2} - \alpha +\sin(\alpha) \cos(\alpha)\Bigr] H,$ as derived in~\cite{donkov:2009a}. As expected, the net force $F_x$ vanishes for equal temperatures $T_1=T_2.$ Also, if the specular wall segments are replaced by diffuse boundaries, $F_x$ vanishes. This shows that in the free-molecular flow regime a $x$-momentum flux in the gas and the corresponding force on the wall cannot be created by virtue of the wall topography alone. 

For comparison with the numerical calculations, we have considered the ratio $F_x/|F_y|.$ 
Such a dimensionless ratio gives an immediate account of the expected magnitude of the effect, since $F_y$ is the pressure force onto the wall. For example,
when the free-molecular flow regime is approximated by considering a gas in a nanochannel at standard conditions, the expression of Eq.~\eqref{eq:fy^top} is the force resulting from a pressure of one atmosphere.

To numerically solve the Boltzmann equation, Monte Carlo schemes were employed. Either a time-splitting Monte-Carlo method~\cite{Babovsky:1989qy,Neunzert:1995uq} with a hard-sphere collision model or a standard Direct Simulation Monte Carlo (DSMC) method with variable hard sphere collision model~\cite{Bird:1994} was used. Both schemes utilize argon molecules as a gas species.

DSMC is a particle method which tracks a number of simulated particles, each being a statistical representation of a large cluster of real molecules. The molecular motions and intermolecular collisions are decoupled over a small time interval. Particles undergo a free convection step followed by a collision step. The convection is treated deterministically while the intermolecular collisions are treated statistically. A detailed description of the DSMC method can be found in~\cite{Bird:1994}.

In all the simulations the cell size is set to be less than  the mean free path and the time step is chosen so that no molecule can cross two cells within one time step. Molecules are initialized based on the average of the wall temperatures and the prescribed Knudsen number. A gas-wall interaction model to compute the post-collision molecular velocities as specified previously in this article is employed.

\begin{figure}[tb!]
\includegraphics[width=3in]{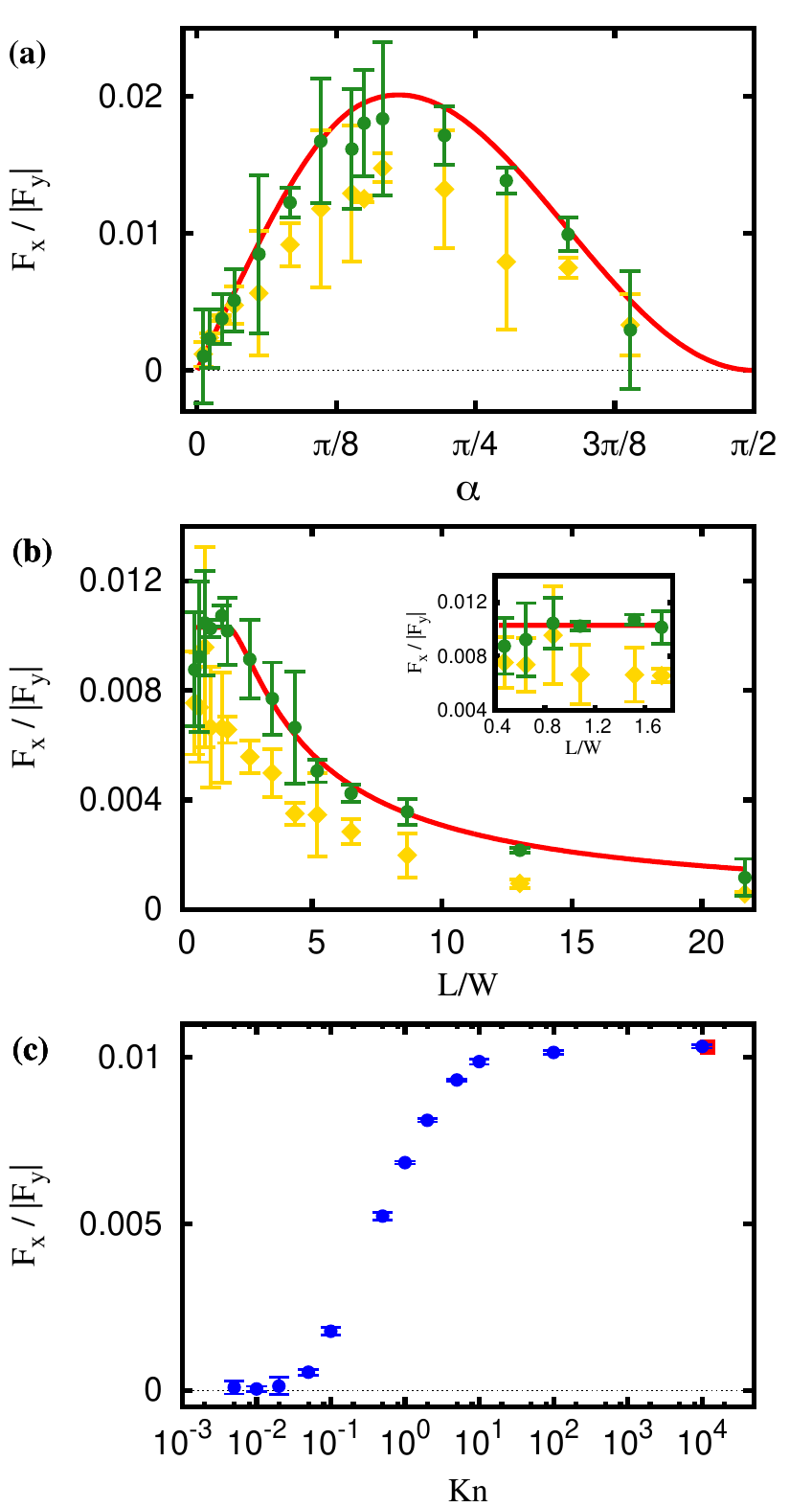}
\caption{\label{fig:ratio-vs-parameter} (Color online) (a) and (b) Analytical (lines) and numerical (points) results for the force ratio as a function of geometric parameters. The circles represent the results obtained for $\text{Kn}=\infty$, the diamonds those for $\text{Kn}=1.$ (a) shows the dependence on  $\alpha$ with $L/W=5/16,$ $H/W = \tan(\alpha)/4.$ (b) displays the dependence on $L/W,$ with $H/W =3/4,$ $\alpha=\pi/3.$ Inset: the data points at small values of $L/W$ in more detail. (c) Numerical results for the force ratio as a function of Knudsen number at $\alpha =\pi/3,H/W=\sqrt{3}/4,L/W=5/16.$ The red square indicates the prediction from the analytical model.}
\end{figure}

\begin{figure}[tbh!]
\includegraphics[width=3in]{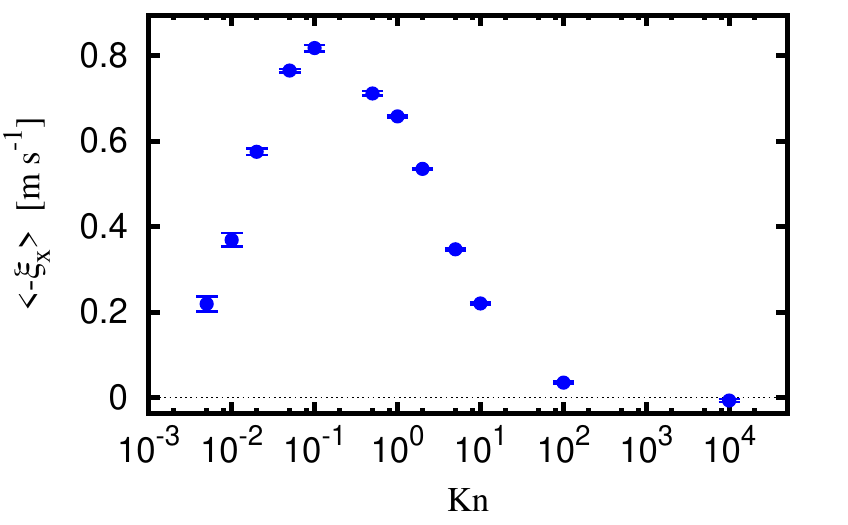}
\caption{\label{fig:flow-rate} (Color online) The area averaged $x$-velocity as a function of the Knudsen number. The averaging is done over a patch connecting the tip of a fin with the upper surface.}
\end{figure}

In the case of the time-splitting method (the data in Figs.~\ref{fig:ratio-vs-parameter}(a-b)), error bars were computed based on three independent runs with $N_{cell}=20,40,80$ particles per computational cell. Following that, a Richardson extrapolation to $N_{cell}=\infty$ was performed, allowing to estimate the error by computing the difference to the $N_{cell}=80$ result. In the case of the standard DSMC method (the data in Figs.~\ref{fig:ratio-vs-parameter}(c)-\ref{fig:flow-pattern}), the error bars represent the maximum deviations computed based on the last one third of total number of samples ranging from 68,000 to 30,000,000. An average of $N_{cell}=40$ particles per cell were used for these simulations.  In all computations the temperatures were fixed to $T_1=300$~K, $T_2=400$~K. The consistency of the two numerical schemes was checked by comparing the results for $F_x$ for specific geometric parameter values at $\text{Kn}=1$ and at $\text{Kn}=\infty.$
The data points obtained with the different methods were found to agree, with deviations within the error bars obtained by the extrapolation, and, at $\text{Kn}=\infty,$ both were within half of their corresponding standard deviation from the analytical prediction, the red square in Fig.~\ref{fig:ratio-vs-parameter}(c).

There are four geometric parameters, $H, L, W, \alpha$, and for the comparison between analytical and numerical calculations we have chosen two ways to fix them, as to have only one independent parameter to be varied. The first one is to fix the distance between the fins $L$ and their base length $H/\tan( \alpha)$ and to change $\alpha,$ the second one is to vary $L$ while keeping the other parameters the same. Figures~\ref{fig:ratio-vs-parameter} (a) and (b) show the analytical and numerical values of $F_x/F_y$ obtained that way. Both $\text{Kn} = 1$ and $\text{Kn} = \infty$ were considered in the numerical studies. In addition, Fig.~\ref{fig:ratio-vs-parameter}(c) shows the dependence of the force ratio on the Knudsen number.

First of all it is worth noting that the analytical curves are well reproduced by the numerical data obtained for $\text{Kn} = \infty$. When the Knudsen number is decreased to one, the data points still follow the same trend, but the magnitude of the dimensionless force is reduced by about $30\%$. We therefore conclude that the analytical solution of the collisionless Boltzmann equation still makes qualitatively correct predictions in the transition flow regime. The asymmetry of the fins (including the effect of the boundary condition) is the origin of the $x$-momentum flux inside the gas and the corresponding force onto the walls. Therefore it is expected that the maximum force values are obtained in a region around $L/W = 1,$ and that the force decreases as the fins become sparser. This is exactly the behavior displayed in Fig.~\ref{fig:ratio-vs-parameter} (b). Fig.~\ref{fig:ratio-vs-parameter} (a) shows that the force is in fact significant: At $\text{Kn} = 1$ maximum values of well above $1\%$ are obtained. As expected, the force vanishes when the gas approaches a state where local thermal equilibrium prevails (cf. Fig.~\ref{fig:ratio-vs-parameter} (c)). The maximum force values are reached in the free molecular flow regime. 

From an energetic point of view, when $F_x$ acts to displace the two parallel surfaces with respect to each other, heat is transformed into mechanical energy, as indicated in the schematic on the right-hand side of Fig.~\ref{fig:device}. We wish to emphasize that this is a novel energy conversion scheme that works without the usual volumetric expansion. Rather than functioning in a cycle, it allows to continuously extract mechanical work from a temperature difference.

\section{Mass flux}

In the limit $\text{Kn} \rightarrow \infty$ where the force becomes maximal, the mass flux in the gas vanishes. Such a scenario bears some analogy with results that have been obtained for closed domains~\cite{Sone:1984uq}. It also shows that the force is not due to the shear stress caused by a fluid flowing over a surface. At $\text{Kn} = \infty$ the phase space distribution of the gas is such that the expectation value of the $x$-momentum flux transferred between the walls is nonzero. The molecules emerging from the lower surface carry a negative $x$-momentum flux, corresponding to the positive value of $F_x$ the surface experiences. Away from the limiting case of infinite Knudsen number, some of these molecules collide with other gas molecules, thereby transferring $x$-momentum to the gas. It is expected that by such a mechanism the gas in the upper portion of the domain into which the fins do not extend is set into motion. This picture suggests that the corresponding mass flux should be opposite to $F_x$.

One can also examine the picture from the other end of the gas regime. In the limit of $\text{Kn} \rightarrow 0$, the local equilibrium conditions ensure that both momentum and mass fluxes are zero and hence no force and mass flow exist. As $\text{Kn}$ increases, non-equilibrium conditions start to prevail. Since the temperature gradient around the fin tips is likely to be the most significant, due to both the sharp tip and the asymmetric boundaries, one would expect that deviations from local thermal equilibrium first occur in these regions, resulting in a larger non-zero momentum flux compared to the gas in other regions. The flow induced by this momentum flux is then expected to have its maximum velocity near the tip and to gradually decrease towards the top wall, resulting in a flow pattern similar to Couette flow. As $\text{Kn}$ increases further, the non-equilibrium region expands into the whole channel. The tip flow is no longer dominant and the flow pattern is dictated by the overall configuration of the channel. 

\begin{figure}[tb!]
\includegraphics[width=3.4in]{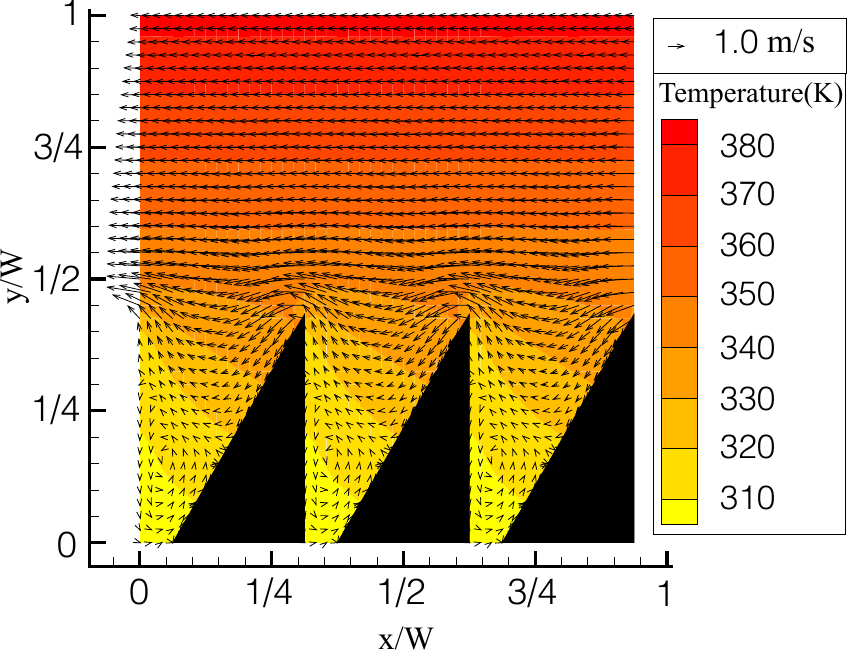}
\caption{\label{fig:flow-pattern} (Color online) The velocity field at $\text{Kn}=0.1$. The background color (grayscale) contours indicate the corresponding temperature field.}
\end{figure}

Figure~\ref{fig:flow-rate} displays numerical results for the average flow velocity as a function of $\text{Kn}$. For these calculations a specific geometry with $\alpha =\pi/3,H/W=\sqrt{3}/4,$ and $L/W=5/16$ was chosen (the same geometry as for the next to last data  point in Fig.~\ref{fig:ratio-vs-parameter}(a)). As expected from the analytical results above and the Navier-Stokes limit of the Boltzmann equation, there is no flow in the limits $\text{Kn} \rightarrow 0$ and $\text{Kn} \rightarrow \infty$. The maximum average flow velocity occurs in the region around $\text{Kn} = 0.1$. The behavior at large Knudsen numbers suggests that the mass flux can be interpreted as a perturbation effect to the zero mass flux, nonzero momentum flux phase space distribution derived in the previous section: The momentum transferred from particle-particle collisions sets the gas into motion. Figure~\ref{fig:flow-pattern} shows the flow pattern obtained at $\text{Kn} = 0.1$. The background color contours indicate the corresponding temperature field. The flow is in the negative $x$-direction and a Couette-like flow pattern with the maximum velocity being near the fin tip is clearly visible, consistent with the simple picture drawn in the previous paragraph. Significant flow velocities of close to $1$~m/s are reached, making the described principle relevant for the pumping of gases.  

\section{Conclusions and outlook} 

We have studied the transport processes in a gas between a flat and a structured surface at different temperatures. We have shown that with suitably chosen surface structures, both a momentum and a mass flux parallel to the surfaces can be induced. The momentum flux takes its maximum values in the free-molecular flow regime, the mass flux (in terms of the average velocity) in the transition flow regime. While it is a priori clear that suitable structures need to break the reflection symmetry, a purely geometrical symmetry breaking based on diffusively reflecting walls is not sufficient. The observed fluxes only occur because a suitable texture comprising surface patches with diffuse and specular reflection conditions have been selected. The described effects may find applications in micro-/nanoscale energy converters and micro-/nanoscale pumping technology.

\acknowledgements
This work was supported by grants No. HA 2696/16-1, KL 1105/17-1 of the German Research Foundation and CERG No. 621408 of the Hong Kong Research Grants Council. 


%

\end{document}